\documentclass[
 reprint,
 superscriptaddress,
 amsmath,amssymb,amsfonts,
 aps,prl,
]{revtex4-2}

\usepackage{graphicx}
\usepackage{dcolumn}
\usepackage{bm}
\usepackage{paralist}

\usepackage{color}
\usepackage{comment}
\usepackage[utf8]{inputenc}
\usepackage[english]{babel}

\def\(({\left(}
\def\)){\right)}
\def\[[{\left[}
\def\]]{\right]}
\newcommand{\s}{\sigma}
\newcommand{\cN}{\mathcal{N}}
\newcommand{\DU}{D_\text{\tiny U}}
\newcommand{\DUFC}{D_\text{\tiny U}^\text{\tiny FC}}

\newcommand{\E}{{\mathbb E}}

\newcommand{\mathsym}[1]{{}}
\newcommand{\unicode}[1]{{}}

\usepackage{graphicx}
\usepackage{bm}
\usepackage{epstopdf}
\usepackage[colorinlistoftodos]{todonotes}

\newcommand{\be}{\begin{equation}}
\newcommand{\ee}{\end{equation}}
\newcommand{\bea}{\begin{eqnarray}}
\newcommand{\eea}{\end{eqnarray}}
\newcommand{\<}{\langle}
\renewcommand{\>}{\rangle}
\newcommand{\prob}{{\mathbb P}}

\newcommand{\f}{\tau}
\def\<{\langle}
\def\>{\rangle}
\newcommand{\sign}{\mathrm{sign}}

\usepackage{subfig}
\usepackage{color}

\usepackage{appendix}

\begin{document}

\title{Unexpected upper critical dimension for spin glass models in a field predicted by the loop expansion around the Bethe solution at zero temperature}

\author{Maria Chiara Angelini}
\affiliation{Dipartimento di Fisica, Sapienza Universit\`a di Roma, P.le Aldo Moro 5, 00185 Rome, Italy}
\affiliation{Istituto Nazionale di Fisica Nucleare, Sezione di Roma I, P.le A. Moro 5, 00185 Rome, Italy}
\author{Carlo Lucibello}
\affiliation{Bocconi Institute for Data Science and Analytics (BIDSA),
  Bocconi University, Via Sarfatti 25, 20100 Milan, Italy}
\author{Giorgio Parisi}
\affiliation{Dipartimento di Fisica, Sapienza Universit\`a di Roma, P.le Aldo Moro 5, 00185 Rome, Italy}
\affiliation{Istituto Nazionale di Fisica Nucleare, Sezione di Roma I, P.le A. Moro 5, 00185 Rome, Italy}
\affiliation{Institute of Nanotechnology (NANOTEC) - CNR, Rome unit, P.le A. Moro 5, 00185 Rome, Italy}
\author{Gianmarco Perrupato}
\affiliation{Dipartimento di Fisica, Sapienza Universit\`a di Roma, P.le Aldo Moro 5, 00185 Rome, Italy}
\author{Federico Ricci-Tersenghi}
\affiliation{Dipartimento di Fisica, Sapienza Universit\`a di Roma, P.le Aldo Moro 5, 00185 Rome, Italy}
\affiliation{Istituto Nazionale di Fisica Nucleare, Sezione di Roma I, P.le A. Moro 5, 00185 Rome, Italy}
\affiliation{Institute of Nanotechnology (NANOTEC) - CNR, Rome unit, P.le A. Moro 5, 00185 Rome, Italy}
\author{Tommaso Rizzo}
\affiliation{Institute of Complex Systems (ISC) - CNR, Rome unit, P.le A. Moro 5, 00185 Rome, Italy}
\affiliation{Dipartimento di Fisica, Sapienza Universit\`a di Roma, P.le Aldo Moro 5, 00185 Rome, Italy}

\date{\today}

\begin{abstract}
The spin-glass transition in a field in finite dimension is analyzed directly at zero temperature using a perturbative loop expansion around the Bethe lattice solution.
The loop expansion is generated by the $M$-layer construction whose first diagrams are evaluated numerically and analytically. The generalized Ginzburg criterion reveals that the upper critical dimension below which mean-field theory fails is $\DU \ge 8$, at variance with the classical result $\DU=6$ yielded by finite-temperature replica field theory.
Our expansion around the Bethe lattice has two crucial differences with respect to the classical one.
The finite connectivity $z$ of the lattice is directly included from the beginning in the Bethe lattice, while in the classical computation the finite connectivity is obtained through an expansion in $1/z$.
Moreover, if one is interested in the zero temperature ($T=0$) transition, one can directly expand around the $T=0$ Bethe transition. The expansion directly at $T=0$ is not possible in the classical framework because the fully connected spin glass does not have a transition at $T=0$, being in the broken phase for any value of the external field.
\end{abstract}

\maketitle

Spin glasses (SG) are the prototype of disordered models.
The fully-connected (FC) mean-field (MF) version, introduced by Sherrington and Kirkpatrick (SK) in \cite{sherrington1975solvable}, was solved forty years ago \cite{parisi1980sequence,*parisi1980order}.
The SK model in a field $h$ undergoes a phase transition from a paramagnetic to a SG phase along the de Almeida-Thouless (dAT) line $h_c(T)$ \cite{de1978stability}, that diverges for $T\to0$. At $T=0$ the SK model is in the SG phase, no matter how strong the external field is.

The solution to the SK model requires the introduction of replicas \cite{parisi1980sequence,*parisi1980order}.
To identify the dAT line one can compute the fluctuations around the paramagnetic solution, via the study of the spectrum of the Hessian of the replicated free energy \cite{de1978stability}.
One can identify three sectors of Hessian eigenvectors, that are called replicon, longitudinal and anomalous \cite{bray1978replica,bray1979replica}.
On the dAT line, the replicon eigenvalue becomes critical and stays critical in the whole SG phase, which is thus marginally stable.
Below the dAT line the replica symmetry is spontaneously broken and the SK model has an exponential number of pure states, organized in an ultrametric structure.
This highly non-trivial solution has been proved to be rigorously exact \cite{talagrand2003generalized,panchenko2013sherrington}.

Beyond MF, things are much less clear. In particular, it is not  known whether the finite-dimensional model with external field has a transition to a SG phase.
Numerical simulations suggest a positive answer for $D=4$ \cite{banos2012thermodynamic}, but for $D=3$ the results are inconclusive due to huge finite-size effects and very large equilibration times \cite{baity2014three,baity2014dynamical}: at the state of the art, it is impossible to decide if a transition exists just based on numerical results. 

Usually, in statistical mechanics, the finite-dimensional behavior of models can be deduced using the powerful method of Renormalization Group (RG) \cite{amit2005field}. 
One can set up a field theory for the order parameter associated with the desired transition, constructing a Lagrangian that is the most general one
compatible with the symmetries of the problem. The basic approximation gives the so-called Landau-Ginzburg (LG) theory.
It corresponds to the assumption that there are no fluctuations in the field and it is exact for the MF-FC model.
The next step is to see how the fluctuations, associated with the short-range interactions, modify the MF picture. Performing this task perturbatively, leads to a \emph{loop-expansion} around the LG solution. Looking at when the one-loop correction becomes important, one identifies the upper critical dimension $\DU$ at which the MF theory does not predict the correct critical behavior anymore: this is the so-called generalized \emph{Ginzburg criterium}. 
At this point, a perturbative expansion around the MF solution can be constructed, with a small parameter $\epsilon=\DU -D$, to see how the MF transitions are modified at dimension $D$ below $\DU$.

Unfortunately, this program cannot be carried out so simply for SG models in a field. The MF theory in the high-temperature phase and the first-order perturbative expansion around it were analyzed in different papers 
\cite{bray1980renormalisation,temesvari2002generic,pimentel2002spin,moore2011disappearance,parisi2012replica,temesvari2017physical}.
Let us stress that the Lagrangian is very complicated: three bare masses, associated with the three sectors,
and eight cubic vertices involving the replica fields.
Forty years of work were not enough to understand the fate of the SG transition in finite dimension.
For $D>\DUFC=6$, the MF-FC Fixed Point (FP) is stable, however, its basin of attraction shrinks to zero approaching $\DUFC$ from above.
The main problem is the absence of a perturbative stable FP below $D=6$ \cite{bray1980renormalisation, pimentel2002spin}: this lack is not a proof of non-existence of SG phase in low dimensions and many scenarios have been put forward. 
Some authors have tried to extract information from the perturbative analysis nonetheless \cite{parisi2012replica,temesvari2017physical}, possibly including quartic interactions \cite{holler2020one} that are known to have a non-trivial role \cite{fisher1985scaling}. 
It could also be possible that a non-perturbative FP exists \footnote{In principle a non-perturbative FP could be important also for $D>\DUFC$ because of the finite basin of attraction of the Gaussian FP.}. 
Recently, the perturbative expansion was computed up to the second-order \cite{charbonneau2017nontrivial,charbonneau2019morphology}, finding a strong-coupling FP that could in principle be stable at any dimension, even above $\DUFC$.
This new FP is in a way ``non-perturbative'' as it cannot be reached continuously from the MF-FC one just lowering the dimension. However, the perturbative analysis in the strong-coupling regime is uncontrolled: thus the existence and relevance of this new FP cannot be stated just with the methods of Refs.~\cite{charbonneau2017nontrivial,charbonneau2019morphology}.

Alternatively, the use of real-space RG methods is the natural choice if we are looking for non-perturbative FP in finite dimensions.
The Ensemble RG (ERG)\cite{angelini2013ensemble} and the Migdal-Kadanoff (MK) RG \cite{angelini2015spin} were applied
to the SG in a field: for high enough dimensions ($D\gtrsim8$) a critical FP at $T=0$ was found, different from the MF-FC one.
We remind that in the FC SK model there is no transition at $T=0$ due to the diverging connectivity, an unrealistic feature that is not present in finite-dimensional models.
However, the MK and the ERG flows are obtained after some crude approximations, as usually done when using non-perturbative RG, that are not exact. Thus they can provide useful indications, but cannot offer a definite answer to the problem.

Recently, a new loop expansion around the MF Bethe solution has been proposed in \cite{altieri2017loop}. 
SG models in a field can be solved on the Bethe lattice (BL) and the finite connectivity allows for local fluctuations of the order parameter.
This is an important feature shared with finite-dimensional systems.
The loop expansion around the Bethe solution is obtained via the $M$-layer construction \cite{altieri2017loop}. One introduces $M$ copies of the original finite-dimensional lattice and generates a new lattice through a local random rewiring of the links. For large $M$ the resulting $M$-layer lattice looks locally like a BL (and thus all observables tend to their MF BL values with small $1/M$ corrections), while at large distances the lattice retains its finite-dimensional character. This has important consequences for critical behavior: close to the MF critical point the system displays MF critical behavior until the correlation length reaches a size where the finite-dimensional nature of the model is dominant and the correct non-MF exponents are observed due to universality.
The $1/M$ expansion (for $M=1$ one recovers the original model) takes the form of a diagrammatic loops expansion with appropriate rules \cite{altieri2017loop} and it is very useful to study critical phenomena. Similarly to field-theoretical loops expansion, one can apply the Ginzburg criterion and identify the upper critical dimension $\DU$ where the corrections alter the MF behavior.
For $D<\DU$ the expansion can then be used to obtain the critical exponents through standard RG treatments. 

The expansion around the BL solution has the same advantages as standard field-theoretical loop expansions, but has a larger range of applicability, as it can be used for any problem that displays a continuous phase transition on the BL. Moreover, while in the classical expansion finite connectivity $z$ is obtained as a result of a $1/z$ expansion around infinite connectivity, in the expansion around Bethe lattice $z$ is finite and fixed from the beginning, introducing less artifacts.
Recent applications of the BL expansion include the Random Field Ising model (RFIM) at zero temperature \cite{angelini2020loop}, the bootstrap percolation \cite{rizzo2019fate} and the glass crossover \cite{rizzo2020solvable}.
It has also been applied to the SG in a field in the limit of high connectivity for $T>0$ \cite{angelini2018one}, showing that in such a limit the expansion is completely equivalent to the standard expansion around the MF-FC solution \cite{bray1980renormalisation,pimentel2002spin}. This is in agreement with the fact discussed in Ref.~\cite{altieri2017loop} that the $1/M$ expansion and the standard field theoretical expansion are completely equivalent if the physics of the model on the BL is like the one on the FC lattice.

In this paper, we study the $M$-layer BL expansion of the SG in a field directly at $T=0$ from both the paramagnetic and the SG phase. 
We are particularly interested in doing the computation directly at $T=0$ because we know that, in this particular situation, degeneracy of eigenvalues can lead to different physics with respect to finite temperature, as happens for example for the RFIM. However, the classical expansion cannot be used directly at $T=0$, because there is no transition in the SK model at $T=0$: the system is in the broken-phase for any value of the field. Things are different in the Bethe lattice, for which the dAT line ends at a finite critical field $h_c$ at $T=0$. Direct expansion around this point is thus possible.  
We show that finite connectivity and zero temperature lead to a critical behavior different from the one of the replicated field theory expansion at finite temperature. In particular, the generalized Ginzburg criterion leads to an upper critical dimension $\DU\ge8$.

To be concrete we consider the model Hamiltonian
\begin{equation}
H = - \sum_{(ij)\in E} J_{ij} \sigma_i\sigma_j - h \sum_i \sigma_i\;,
\end{equation}
where the spins take the values $\sigma_i=\pm 1$, $h$ is a constant external field
\footnote{The physics of the model does not change if the external field is replaced by a random field.} and the quenched couplings $J_{ij}$ have a Gaussian distribution with $\overline{J}=0$,
$\overline{J^2}=\frac{1}{z-1}$, $z$ being the (fixed) connectivity 
of the model.
The first sum is over the set of edges $E$ of a $D$-dimensional lattice.

Approaching the transition from the paramagnetic side the order parameter is zero and we analyze, as usual, the behavior of spin correlations. 
Following Ref.~\cite{altieri2017loop}, a generic correlation or response function $G(x)$ between two points at distance $x$ on the original lattice is given at leading order in $1/M$ by
\begin{equation}
G(x)=\frac{1}{M}\sum_{L=1}^\infty{\cal N}(x,L)\, G^{\text{\tiny{BL}}}(L)\;,
\label{eq:G(x)}
\end{equation}
where ${\cal N}(x,L)$ is the number of non-backtracking paths of length $L$ connecting the two points at distance $x$ on the  original lattice ($M=1$) and $G^{\text{\tiny{BL}}}(L)$ is the analyzed correlation function between two spins at distance $L$ on a BL with connectivity $z=2D$. 
While ${\cal N}(x,L)$ is known  \cite{altieri2017loop}
\begin{equation}
{\cal N}(x,L) \propto (2D-1)^L \exp\left(-x^2/(4L)\right) L^{-D/2},
\label{eq:N_L}
\end{equation} 
the crucial model-dependent quantity to be computed is $G^{\text{\tiny{BL}}}(L)$.
Working at $T=0$ it is worth focusing on the response function $R_{ij}$ defined for the spin-glass model via the following procedure: being $\bm\sigma^\star$ the ground state (GS) configuration; compute the new GS under the constraint $\sigma_i=-\sigma_i^\star$; if also $\sigma_j$ flips, then $R_{ij}=1$, otherwise  $R_{ij}=0$.
One can show \footnote{For more details see Supplemental Material [SM], which includes Refs. \cite{Brezin1976, vontobel2013counting, angelini2017real}} that the average response function on the BL can be computed exactly by applying $L$ times  an integral operator. Consequently, its behavior at large $L$ is given by
\begin{equation}
R^{\text{\tiny{BL}}}(L)\propto \lambda^L\;,
\label{eq:R_L}
\end{equation}
where $\lambda$ is the largest eigenvalue of the integral operator (more details on $\lambda$ in the SM). It goes to $\lambda_c=\frac{1}{2D-1}$ at the critical point of the BL, such that the total response diverges and the paramagnetic solution is no longer stable \footnote{The integral operator is the one in Eq.~(26) of Ref.~\cite{parisi2014diluted}. The connection with $R_{ij}$ follows from the fact that spin $\sigma_j$ flips after a flip of $\sigma_i$ only if an infinitesimal change of the field acting on site $i$ propagates to site $j$ and this condition is exactly enforced by the integral operator.}.
Inserting Eqs.~(\ref{eq:R_L}) and (\ref{eq:N_L}) into Eq.~(\ref{eq:G(x)}), we obtain for the Fourier transform of the response function in the small momentum region:
\begin{align}
\nonumber
R(p)&\propto \frac{1}{M}\sum_{L=1,\infty} \left[\lambda\cdot(2D-1)\right]^L \exp(-L \, p^2)\\
&\simeq  \frac{1}{M} \int_0^\infty dL\ \exp\left(-L (p^2 +\tau)\right)= \frac{1}{M} \,\frac{1}{p^2+\tau}
\label{eq:R(p)}
\end{align}
with $\tau\equiv -\log(\lambda(2D-1))$. Note that $\tau\to0$ when $\lambda\to\lambda_c$: at leading order the response  has the form of the bare propagator in a field theory and becomes critical at the BL critical point.

Let us now look at the $1/M^2$ correction to the bare propagator. According to Ref.~\cite{altieri2017loop}, this is given by the sum of the contributions coming from all the paths that connect the two points on the original lattice containing just one topological loop.
The contribution of a specific topological diagram in Fourier space is
\begin{equation}
\widetilde{G}_{\text{loop}}(p)=\frac{1}{M^2}\sum_{\vec L} \ \cN(p,{\vec L}) G^{\text{\tiny{BL}}}_{\text{loop}}(\vec L)\;,
\label{eq:Gloop}
\end{equation}
where $\vec L$ is a vector containing the lengths of each line in the topological diagram and the factor $\cN(p,\vec L)$ accounts for the number of such topological diagrams on the original regular lattice with $M=1$.
The term $G^{\text{\tiny{BL}}}_{\text{loop}}(\vec L)$ is again the only term depending on the model:
it is the so-called \emph{line-connected} value \cite{altieri2017loop} that the observable takes on a BL in which the analyzed topological loop has been manually inserted.
The term ``line connected'' means that one should add the value of the observable evaluated on each of the subgraphs that are obtained from the original structure 
by sequentially removing its lines times a factor $-1$ for each line removed.

Let us point out two crucial differences between this expansion and the standard expansion around LG theory: 
\begin{itemize}
\item the latter has just cubic vertices, while in the BL expansion vertices of all degrees can be present;
\item the diagrams of the BL expansion have a clear physical meaning while the Feynmann diagrams of the standard expansion are just a smart way to compute the desired corrections.
\end{itemize}

\begin{figure}[t]
\begin{center}
\includegraphics[width=\columnwidth]{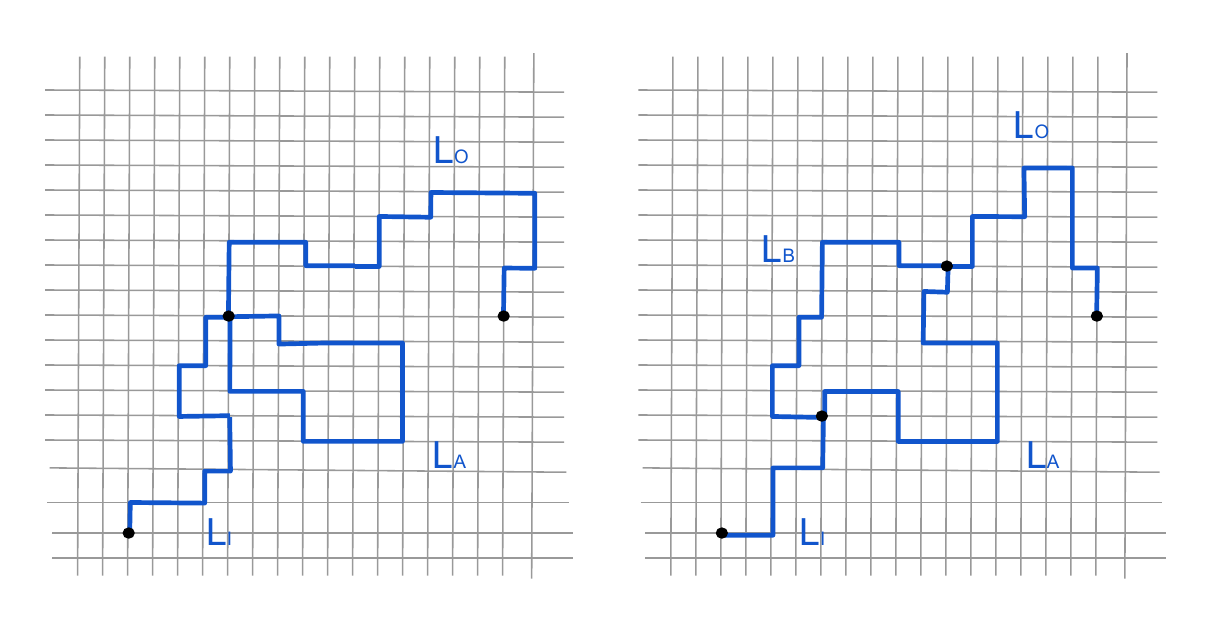}
\caption{One loop topological diagrams relevant for the first order correction around the BL: the ``quartic loop'' on the left has a vertex with four lines, while the ``cubic loop'' on the right has only vertices with three lines.}
\label{Fig:loop}
\end{center}
\end{figure}

At one loop we consider the two diagrams shown in Fig.~\ref{Fig:loop}. 
The left one has a quartic vertex, for this reason it is not included in the standard cubic theory. 
We compute $G^{\text{\tiny{BL}}}_{\text{loop}}(\vec L)$ on this diagram with the same tools as for the 0-loop term (all the details in the SM).
The resulting contribution to the response function coming from this quartic loop is $R^{\text{\tiny{BL}}}_{\text{4-loop}}(\vec L)\propto L_A\lambda^{\Sigma(\vec L)}$ , where $\Sigma(\vec L)$ is the sum of all $L$'s, i.e.\ $\Sigma(\vec L)=L_A+L_I+L_O$ in this diagram, and $\lambda$ is the same eigenvalue on the BL as in the previous discussion. 
The cubic loop (on the right in Fig.~\ref{Fig:loop}) has cubic vertices and is already present in the LG theory. 
Its behavior should be analyzed when $L_A$ and $L_B$ are large, because we checked that when one of the two internal legs is short, the diagram reduces to the quartic loop.
For large $L_A$ and $L_B$, we obtain $R^{\text{\tiny{BL}}}_{\text{3-loop}}(\vec L)\propto \frac{L_AL_B}{L_A+L_B}\lambda^{\Sigma(\vec L)}$, with $\Sigma(\vec L)=L_A+L_B+L_I+L_O$.

The term $\cN(p,\vec L)$ has already been computed \cite{angelini2020loop} and it reads respectively for the quartic and cubic loops
\begin{eqnarray}
\cN(p,\vec L)&\propto&\frac{(2D-1)^{\Sigma(\vec L)}}{{L_A}^{D/2}} e^{-\left(L_I +L_O\right)p^2}\;,\\
\label{eq:Ntadpole}
\cN(p,\vec L)&\propto&\frac{(2D-1)^{\Sigma(\vec L)}}{\left(L_A+L_B\right)^{D/2}}
e^{-\left(L_I +L_O+\frac{L_AL_B}{{L_A+L_B}}\right)p^2}\;. \label{eq:Nphi3}
\end{eqnarray}
Inserting the above expression and $R^{\text{\tiny{BL}}}_{\text{3-loop}}(\vec L)$ in Eq.~(\ref{eq:Gloop}), we obtain the correction to the response given by the cubic loop.
In order to apply the Ginzburg criterion, it is more convenient to consider the inverse susceptibility
\begin{multline*}
(M R(p))^{-1} = \tau+p^2+\\
+\frac{c}{M} \sum_{L_A,L_B}  \frac{L_AL_B}{(L_A+L_B)^{D/2+1}} e^{-L_A \tau-L_B \tau-\frac{L_AL_B}{L_A+L_B} p^2}\;,
\end{multline*}
that can be rewritten as 
\begin{eqnarray}
(M R(p))^{-1} & = & A \, (\tau-\tau_c)+ B \, p^2 + O(p^4)\;,\quad\text{with}\nonumber\\
\tau_c & = & \frac{c}{M} \sum_{L_A,L_B}  \frac{L_AL_B}{(L_A+L_B)^{D/2+1}}\;,
\\
A & = & 1 - \frac{c}{M} \sum_{L_A,L_B}  \frac{L_AL_B}{(L_A+L_B)^{D/2}}\;,
\\
B & = & 1 - \frac{c}{M} \sum_{L_A,L_B}  \frac{L_A^2L_B^2}{(L_A+L_B)^{D/2+2}}\;.
\end{eqnarray}
We see that for large but finite $M$, the $M$-layer lattice has the same critical behavior of the BL  ($M=\infty$), with small $O(1/M)$ shifts of the critical temperature and of the constants $A$ and $B$. However, the above sums over $L_A$ and $L_B$ are divergent respectively for $D \leq 6$, $D \leq 8$ and $D \leq 8$ and thus the Ginzburg criterion tells us that the critical exponents cannot be those of the Gaussian theory below $D=8$.
The same argument applied to the quartic loop would give a critical dimension equal to 6 (the diagram indeed appears in the computation of the connected correlation of the RFIM \cite{angelini2020loop}) and allows to neglect the quartic loop with respect to the cubic one.
We also checked that the generalized Ginzburg criterion coming from the Replica Symmetry Breaking (RSB) phase predicts an upper critical dimension $\DU\ge8$, in perfect agreement with the computation in the symmetric phase \cite{unpublished}.

To go below the upper critical dimension we rescale lengths as  $L=x/\tau$ and momenta as $p^2=k^2\tau$, obtaining
\begin{multline}
(M R(p))^{-1}/ \tau  = 1 +k^2 +\frac{c \, \tau^{D/2-4}}{M} \times\\
\times  \int_{\tau/\Lambda}^{\infty} dx_A\,\int_{\tau/\Lambda}^{\infty} dx_B  \frac{x_A\,x_B\, e^{-x_A -x_B -\frac{x_A x_B}{x_A+x_B} k^2}}{(x_A+x_B)^{D/2+1}}\,.
\end{multline}
The above expression shows that loop corrections are not negligible for $D<8$ when $\tau\to 0$.
Indeed for $D<8$ the integral would be divergent at short distances if not for the lattice cutoff $\Lambda$. One should check if, by standard mass, field, and coupling constant renormalization, the above 1-loop diagrams and higher-order diagrams as well can be made finite in the limit $\Lambda \to \infty$. Then the critical exponents can be computed by standard methods \cite{parisi1988statistical,zinn2002quantum,le1991quantum} provided an $O(\epsilon)$ non-trivial FP of the $\beta$ function can be identified (at variance with the $T>0$ case \cite{bray1980renormalisation}): this program is currently underway.
An interesting question is if this putative zero-temperature FP describes also the $T>0$ physics, i.e.\ if the temperature is an irrelevant operator in the Wilson RG sense.
We already mentioned that the expansion around the BL was applied to the SG in a field for $T>0$ and in the limit of large $z$ in Ref.~\cite{angelini2018one}. 
Even if we take the limit $T\to0$ of that expansion, the 1-loop correction results to be of the standard form (the detailed computation is in the SM). Finite connectivity is thus a crucial ingredient in the computation, and the limits $z\to\infty$ and $T\to0$ cannot be exchanged. This is a clear indication that for SG models the expansion around the FC model cannot describe the behavior of finite-dimensional systems.

We emphasize that the FP we have found in this work by expanding around the BL is different from the finite temperature MF-FC one even for $D>8$.
Indeed when $T>0$ one can demonstrate that the critical behavior of all the possible correlation functions is the same 
(mainly because they all receive a critical contribution by the only critical eigenvalue, the replicon \cite{de2006random,angelini2018one}).
However, if the relevant FP is a $T=0$ one, different correlation functions could decay differently (this effect is linked to the degeneracy of the three eigenvalues that become all critical when $T\to0$),
so one should look at them all. 
This is what happens in the RFIM, whose physics is governed by a $T=0$ FP and whose correlation function associated with disorder fluctuations decays more slowly than the one associated with thermal fluctuations \cite{de2006random}.
The same behavior is predicted by the MK RG of Ref.~\cite{angelini2015spin} for the SG in a field.
We leave the analysis of the disorder correlation function to future work.

We just looked to the first order correction in the BL expansion. Going beyond this computation is really hard: the second-order terms are already much involved in the standard expansion (see ref. \cite{charbonneau2017nontrivial}), in the BL expansion one should consider in addition also diagrams with quartic vertices. A simple evaluation of the diagrams with power counting method is not possible because exact cancellation could happen and a quantitative computation is needed. 
The possibility that two-loop diagrams (or even higher order diagrams) diverge at a dimension $D>8$ cannot be excluded, even if it would be quite unexpected, having never been observed in any known model. For this reason the best that we could say is that $\DU\ge8$.
The identification of simple rules for the computation of Feynmann diagrams is one of our planned next steps.

A final remark on the value $\DU=8$: in Ref.~\cite{angelini2015spin} the upper critical dimension was found to be $D\simeq 8$ with the MK RG method, while for $D<8$ no stable SG phase was found.
It is thus important to numerically address the problem of the identification of $\DU$. Unfortunately,
numerical simulations cannot be performed directly on hyper-cubic lattices of such high $D$. For this reason the perfect candidates are the one-dimensional Long-Range (LR) models. There exist two versions of these:
\begin{compactitem}
    \item[--] a fully connected version, where the variance of the couplings between any two spins decays with their mutual distance $r$ as a power-law $r^{-\sigma}$ \cite{kotliar1983one,katzgraber2003monte, katzgraber2003geometry, katzgraber2005probing, katzgraber2008spin}. When $\sigma\to0$, one recovers the MF SK model. 
    \item[--] a diluted version, introduced to reduce the simulation time, where all couplings are $O(1)$ and present with a probability decaying as $r^{-\sigma}$ \cite{leuzzi2008dilute,katzgraber2009study,leuzzi2009ising, larson2013spin}. When $\sigma\to0$, one recovers the BL model. 
\end{compactitem}
These models are proxies for short range (SR) models in higher dimensions: in both cases changing $\sigma$ is equivalent to change the dimension $d$ of the corresponding SR model. Relations 
that link $\sigma$ and $d$ have been studied in detail \cite{banos2012correspondence,Angelini2014Relations}.
The numerical investigation of LR models was focused on the existence of a transition for the SG in external field below $\DU$, for effective SR dimensions $d\simeq3,4,5$ \cite{leuzzi2009ising,katzgraber2009study,banos2012correspondence}, while only data in $d\simeq10,20$ were collected in the assumed MF region \cite{leuzzi2009ising,katzgraber2009study,aspelmeier2016finite}.
One should then check which is $\DU$ with and without field: Is there a signature in LR models that $\DU$ changes from 6 to 8 when an external field is added? One could look at the critical exponent $\nu$ as a function of $d$ that should have a kink exactly at $\DU$ in LR models \cite{angelini2013ensemble}.
Moreover, one could look at which kind of finite size scaling (MF or non-MF) leads to a better collapse of numerical data at different sizes, depending on $d$ \cite{aspelmeier2016finite}. 
We expect that $\DU$ is different for the FC and the diluted version of the LR models. In the past, the given explanation for their equivalence was based on standard field theoretical analysis \cite{kotliar1983one}. However, if the BL-expansion gives different prediction w.r.t. the FC standard expansion, FC and diluted version of LR models should display differences, because of their different limits when $\sigma\to 0$.

This research has been supported by the European Research Council under the European Union Horizon2020 research and innovation program (grant No.~694925 -- Lotglassy, G.~Parisi).

\bibliography{arxiv}

\onecolumngrid

\vspace{5mm}
\begin{center}
\bf \large{Supplemental Material for ``Unexpected upper critical dimension for spin glass models in a field predicted by the loop expansion around the Bethe solution at zero temperature''}
\end{center}

\section{The $M$-layer expansion around the fully-connected lattice and around the Bethe Lattice: differences and analogies}

In this section we will show how to construct a fully-connected $M$-layer model, that has been shown to be equivalent to the standard Ginzburg-Landau theory associated to the given original model \cite{Brezin1976}. We will then show how to analogously build the Bethe-lattice $M$-layer construction, following ref. \cite{altieri2017loop}.
For simplicity, let us consider a standard Ising model on a finite dimensional lattice:
\begin{equation}
H=-\frac{1}{2}\sum_{(i,j)} J\, \sigma_i\,\sigma_j,
\label{eq:Ising}
\end{equation}
where we indicate with $(i,j)$ the couples of nearest neighboring spins.
At this point we create a fully-connected $M$-layer lattice in this way:
On each site of the lattice, we put a stack of $M$ Ising spins. Then we couple each spin ferromagnetically with the $M$ spins on each of its $2D$ nearest neighbors. In the $M\to\infty$ limit, the connectivity of each single spin diverges. Therefore we rescale the couplings by a factor $M$, to have a good $M\rightarrow \infty$ limit. We denote with $\sigma_i^\alpha$ the configuration of spin $i$ in the layer $\alpha$, 
and write the whole partition function as
\begin{equation}
Z_M = \sum_{\{\sigma^\alpha_i\}} \exp\left\{\frac{\beta}{2M}\sum_{(i,j)} J\sum_{\alpha,\alpha'}^M \sigma^\alpha_i\sigma^{\alpha'}_j\right\},
\end{equation}
which, performing a standard Hubbard-Stratonovich transformation, becomes:
\begin{equation}
Z_M \propto \int_{-\infty}^{+\infty} \prod_{i=1}^{N} \text{d} m_i\  \exp{  \left \lbrace M \left[ -\frac{\beta}{2} \sum_{ij} m_i (J^{-1})_{ij} m_j +\sum_i \log {\left[ 2 \cosh(\beta m_i ) \right] } \right] \right \rbrace }.
\label{Zm-fc}
\end{equation}
For $M=1$ we recover the original system. Performing a saddle point evaluation of the above integral one obtains exactly the mean field (fully-connected) solution, that can thus be interpreted as the limit of large $M$ of the fully-connected $M$-layer construction. To go beyond mean-field one can perform a $1/M$ expansion: {\it The relevant terms of this expansion are the same that would be obtained by the loop expansion of the Ginzburg-Landau theory}, as shown in detail in ref. \cite{altieri2017loop}. 

Taking inspiration from the fully-connected construction, one can also construct an $M$-layer expansion around the Bethe lattice.
Starting, for simplicity, from the same model in eq. (\ref{eq:Ising}), 
again we create $M$ layers of the original lattice. Then for each edge $(i,j)$ in the original lattice, we create links among the $M$ layers choosing one of the $M!$ possible matchings among the $M$ copies of $i$ and $j$. 
In this construction, the connectivity of the $M$-layer model remains that of the original model.
The partition function of the whole system thus reads:
\begin{equation}
Z_M^{\pi}= \sum_{\{\sigma^\alpha_i\}} \exp\left\{\frac{\beta}{2M}J\sum_{(i,j)}\sum_{\alpha=1}^M\sigma_i^{\alpha}\sigma_j^{\pi^{i,j}_{\alpha}}\right\};
\end{equation}
with $\pi$ a permutation of the links independently chosen for each couple $(i,j)$ (averages over the possible $\pi$ should then be taken).
Analogously to the fully connected case, one can compute the average free-energy in the $M\to\infty$ limit performing a saddle point computation, obtaining exactly the Bethe solution of the model \cite{vontobel2013counting}. This is expected once we observe that in the $M\to\infty$ limit, the $M$-layer system has no loop of finite length: it is locally a tree, for which the Bethe approximation should be exact.

In ref. \cite{altieri2017loop} the $1/M$ expansion around the Bethe lattice solution is analyzed, analogously to what is done for the fully connected model. However, the expansion around the Bethe lattice introduces less artifact w.r.t. the standard expansion, being the connectivity finite and fixed to the one of the original system from the beginning.

When quenched disorder is present in the Hamiltonian, one should first generate
an instance of the $M$-layer lattice and then assign the $J^{\alpha\beta}_{ij}$ as independent random variables. Within the replica formalism often used to deals with such systems, this corresponds to first introduce replicas in the original system and
average over disorder, in this way the resulting system will be homogeneous, and then apply the $M$-layer construction. 

When the fully-connected MF solution and the Bethe lattice solution of the original model are of the same type, the Bethe lattice $M$-layer expansion gives exactly the same results as the standard expansion. This is shown for example for the Ising model in ref. \cite{altieri2017loop}, but also for the spin-glass model in a field at $T>0$ in the limit of large connectivity in ref. \cite{angelini2018one} where the highly non-trivial Bray-Roberts results are exactly found \cite{bray1980renormalisation}.
Conversely, when the nature of the transition in the Bethe lattice is different from the fully-connected one, the two expansions lead to different results, this is the case for the Random Field Ising Model \cite{angelini2020loop}.

\section{Spin glass models on the Bethe Lattice}
We consider a spin glass model of $N$ Ising spins, $\sigma_i=\pm1$,  with Hamiltonian
$$
\mathcal{H} = - \sum_{(ij)\in E} J_{ij} \sigma_i\sigma_j - H \sum_i \sigma_i\;.
$$
The edge set $E$ defines the interaction graph, which is a random regular graph of fixed degree $z$, also known as Bethe Lattice (BL), $H$ is a constant external field and the quenched couplings $J_{ij}$ are random variables extracted from a Gaussian distribution with $\E_J[J]=0$ and  $\E_J[J^2]=\frac{1}{z-1}$ (this scaling ensures a well-defined Hamiltonian in the $z\to\infty$ limit).

A complete description of this model, even at $T=0$, can be found in Ref.~\cite{parisi2014diluted}, however, we report here some of the main ingredients useful for the subsequent computations.
To solve the model in the high-temperature region, we consider cavity fields $h_{i \to j}$ and $u_{i\to j}$ defined on each edge of the graph.
They parametrize, respectively, the marginal probability distribution on $\s_i$ in the cavity graph where edge $(ij)$ has been removed, 
and the marginal probability distribution on $\s_j$ just considering the information coming from the edge $(ij)$, 
in other words the marginal probability in the cavity graph where all edges involving vertex $j$, but $(ij)$, have been removed.

The BL has the special property that in the large $N$ limit the loops of finite length have a vanishing density. In other words, the BL is locally tree-like. For this reason, the different cavity fields $u_{i\to j}$ arriving in $j$ 
from its neighbors can be considered as independent in the large $N$ limit: this property makes the BL a mean-field solvable model.
In fact, one can write self-consistent equations involving the cavity fields that at $T=0$ read
\begin{subequations}
\label{eq:BP}
\begin{eqnarray}
h_{i\to j} &=& H + \sum_{k \in \partial i \setminus j} u_{k\to i}
\label{eq:BP1}\\
u_{i\to j} &=& \sign(h_{i\to j} J_{ij})\;\min(|h_{i\to j}|,|J_{ij}|)
\label{eq:BP2}
\end{eqnarray}
\end{subequations}
where $\partial i$ is the set of neighbors of $i$.
These equations allow us to solve the model on a given (locally tree-like) graph. However, in the large $N$ limit, if we are interested in computing a self-averaging observable, like a free-energy or a correlation function, it is enough to known the probability distribution of the cavity field that satisfies the following self-consistency equation
\begin{equation}
\label{eq:P}
P_B(u) = \E_J \int \prod_{i=1}^{z-1} P_B(u_i) du_i\,\delta\left(u - \sign\Big(J(H+\sum_i u_i)\Big) \min\Big(|J|,|H+\sum_i u_i|\Big)\right)
\end{equation}
This distribution gives the correct statistical description for the cavity messages that in turn provide the correct marginal probabilities in the paramagnetic phase of a spin glass model defined on a very large $z$-regular random graph (a BL).

In the main text, an expansion around the BL for spin glass models on finite-dimensional lattices was used to compute the critical behavior of the two-point connected correlation function. 
Due to the symmetry of the coupling distribution, the first non-trivial two spins correlations for the SG in a field are the squared correlations. 
In particular, the connected squared correlation between two
points at distance $x$ is defined as
\begin{equation}
G_c(x)=\E_J\left[\<\sigma_0 \sigma_x\>^2-\<\sigma_0\>^2 \<\sigma_x\>^2\right]\;.
\label{eq:Gc}
\end{equation}
where $\<\cdot\>$ denotes the thermal average while $\E_J$ is the average over the quenched disordered couplings, as before. $G_c(x)$ is the correlation function associated to the thermal fluctuations
and it goes to 0 when $T\to0$ as $G_c(x)=O(T^2)$. For this reason, in the following we will define a rescaled connected correlation function that stays finite at $T=0$.
The associated susceptibility is the so-called spin-glass susceptibility and diverges at the dAT line in the MF solution.
While in Ref. \cite{angelini2018one} the computation of $G^{\text{\tiny{BL}}}(L)$, and of its 1-loop correcting term $G^{\text{\tiny{BL}}}_{\text{loop}}(\vec{L})$, 
was done analytically for all the possible two-points squared correlations in the limit $z\to \infty$ and $T>0$ for the SG in a field, 
the analytical computation is unfeasible 
when $z$ is finite. 

To obtain the zero-order expansion, one needs to compute the correlation between two points at distance $L$ on a BL. 
Since on a BL there exists only one path of finite length between two given spins $\sigma_1$ and $\sigma_2$, we can obtain an effective two-spins Hamiltonian by integrating out all the internal spins along the path
\begin{equation}
\mathcal H [\s_1,\s_2] = -h_1 \s_1 -J_{12} \s_1\s_2 -h_2 \s_2,
\label{eq:TwopointsH}
\end{equation}
The effective Hamiltonian is fully determined by a triplet $(h_1,h_2,J_{12})$ of effective fields and effective coupling.
At zero temperature, the Gibbs measure is concentrated on the ground state $(\s^*_1,\s^*_2)$ of the effective Hamiltonian that can be easily computed from Eq.~(\ref{eq:TwopointsH}).
Since we are at $T=0$, the connected correlation function in Eq.~(\ref{eq:Gc}) is ill-defined; therefore, we work with its rescaled version that we called response function \cite{angelini2020loop}
\begin{equation}
    R_{ij} = \mathbb{P}\big[\langle\s_i\rangle_j=-\s^*_i\big]\;,
\end{equation}
where $\langle \cdot\rangle_j$ denotes the expectation over the ground state of the system conditioned to the flipping of the spin $\s_j$, i.e.\ $\langle\s_j\rangle_j = -\s^*_j$. In practice, $R_{ij}=1$ if $\s_i$ flips due to the flipping of $\s_j$, and $R_{ij}=0$ otherwise.

\begin{table}[h]
\centering
\caption{\label{table:GSrules}
Rules for computing the ground state configuration $(\s^*_1,\s^*_2)$ of the Hamiltonian in Eq.~(\ref{eq:TwopointsH}) given the triplet of cavity messages $(h_1,h_2,J_{12})$.}
\begin{tabular}{c|c|c}
& $\s^*_1$& $\s^*_2$ \\
\hline
$|h_1| < \min(|J_{12}|,|h_2|)$ & $\sign(J_{12}h_2)$ & $\sign(h_2)$\\
$|h_2| < \min(|J_{12}|,|h_1|)$ & $\sign(h_1)$ & $\sign(J_{12}h_1)$\\
$|J_{12}| < \min(|h_1|,|h_2|)$ & $\sign(h_1)$ & $\sign(h_2)$\\
\end{tabular}
\end{table}

Making use of the rules in Tab.~\ref{table:GSrules} to compute the ground state of the two-spins effective Hamiltonian, it is quite easy to show that, in terms of the effective triplet $(h_1,h_2,J_{12})$, the response function can be written as
\begin{equation}
R_{12}=\mathbb{P}\big[|J_{12}| > |h_1|\big],
\label{eq:r-Jh}
\end{equation}
The crucial quantities for the computation of the response $R^{\text{\tiny{BL}}}(L)$ between two spins at distance $L$ on a BL are thus the triplets of effective coupling and fields at distance $L$.
We keep track of the distribution of triplets in two different ways: in Sec.~\ref{Sec:num} we explain how to perform unbiased Monte Carlo sampling to propagate an empirical distribution of triplets along a line; in Sec.~\ref{Sec:Ansatz} instead, we will approximate the distribution introducing an analytical Ansatz that we argue to be exact in the large $L$ limit. In both cases, we build on previous approaches 
proposed for the Random Field Ising Model (RFIM) in Ref.~\cite{angelini2020loop}.

\section{Numerical computation of the distribution of effective triplets on a line}\label{Sec:num}

Triplets can be computed in a recursive fashion. Let us join two chains, the first one between $\sigma_1$ and $\tau$, 
characterized by the triplet $(u_1,u_{\tau, 1},J_1)$, and the second one between $\tau$ and $\sigma_2$ identified by $(u_{\tau, 2},u_2,J_2)$. 
In order to compute the triplet describing the effective Hamiltonian between $\s_1$ and $\s_2$ we need to sum over $\tau$ and keep only the lowest energy term because we are working at $T=0$:
\begin{equation}
\begin{aligned}
\mathcal H(\s_1,\s_2) &= -\s_1u_1-\s_2u_2+\min_\f \left[-J_1\s_1 \f - h \f -J_2 \f\s_2 \right]\\
& \equiv E -(u_1+u_1')\s_1 -J_{12} \s_1\s_2-(u_2+u_2')\s_2
\label{eq:H_min}
\end{aligned}
\end{equation}
where $h=H+u_{\tau,1}+u_{\tau,2}+\sum_{k\in \partial \tau \setminus 1,2}u_{k\to \tau}$ is the total field acting on spin $\tau$, while $u_{k\to \tau}$ are $z-2$ independent random variables extracted from $P_B(u)$ and
\begin{align}
u_1'&=(A-B+C-D)/4\;,\nonumber \\
u_2'&=(A-B-C+D)/4\;,\label{eq:rules}\\
J_{12}&=(A+B-C-D)/4\;, \nonumber 
\end{align}
with $A=|J_1+J_2+h|$, $B=|J_1+J_2-h|$, $C=|J_1-J_2+h|$, $D=|-J_1+J_2+h|$.

In practice, we start from a population $P_{L=1}(u_0,u_1,J_1)$ of random triplets $(0,0,J)$, with $J$ extracted from a Gaussian distribution.
We evolve the population $P_{L-1}$ into the population $P_{L}$ following the rules summarized in Eq.~(\ref{eq:rules}), where each triplet of the population $P_{L-1}$ is joined to a 
random triplet $(0,0,J)$ and $z-2$ cavity fields $u_{k\to \tau}$ extracted from $P_B(u)$ are added on the central spin.

Unfortunately, this procedure is very ineffective, because at each step a constant fraction of the population (the one satisfying the condition $|h|>|J_{L-1}|+|J|$) 
produces a new triplet with $J_L=0$: in this way
the part of the population keeping information about branches with non-zero effective couplings shrinks very fast during the iterations.

To amplify this signal, one could evolve two populations of the same size: one population stores the triplets corresponding to the branches with $J_L\neq0$, 
while a second population keeps the pairs $(u_0,u_L)$ along branches with $J_L=0$. This is what has been done in the RFIM case. 
However, for the SG, looking carefully at the evolution rules in eq.(\ref{eq:rules}), one can notice that among the triplets $(u_0,u_L,0)$, the pair of fields could become independent in some situations.
To further amplify the signal then, we evolve three populations of the same size:
the population $\mathcal{A}_L$ stores the triplets along branches with $J_L\neq0$,
the population $\mathcal{B}_L$ keeps correlated pairs $(u_0,u_L)$ along branches with $J_L=0$,
while the population $\mathcal{C}_L$ keeps pairs $(u_0,u_L)$ of independent fields along branches with $J_L=0$. When an effective triplet $T_{L-1}$ at length $L-1$ 
is joined with a triplet $(0,0,J)$ to form a new triplet $T_{L}$, 
we encounter the following different cases:
\begin{itemize}
\item $T_{L-1}\in\mathcal{A}_{L-1}$ and $|h|<|J_{L-1}|+|J|$ $\quad \rightarrow\quad$ $T_L\in\mathcal{A}_L$
\item $T_{L-1}\in\mathcal{A}_{L-1}$ and $|h|>|J_{L-1}|+|J|$ $\quad \rightarrow\quad$ $T_L\in\mathcal{B}_L$
\item $T_{L-1}\in\mathcal{B}_{L-1}$ and $|h|<|J|$ $\quad \rightarrow\quad$ $T_L\in\mathcal{B}_L$.
\end{itemize}
In all the other cases $T_L\in\mathcal{C}_L$.
At the same time, we measure the probabilities $p_L$ and $c_L$ that are the weights of the $\mathcal{A}_L$ and $\mathcal{B}_L$ populations respectively.

\begin{figure}
\centering
\includegraphics[width=0.65\columnwidth]{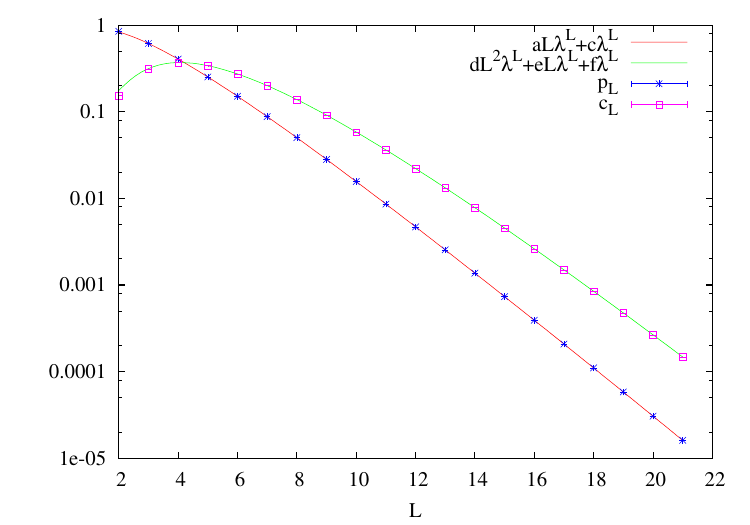}
\caption{\label{Fig:numerical_pL}
Probability $p_L$ to have $J_L\neq0$ and probability $c_L$ to have 
null coupling and correlated fields $(u_0,u_L)$ on a chain of length $L$ in a Bethe lattice: both decay exponentially in $L$.
Measurements are taken at the critical field $H_c=0.358$ for $z=3$. Errors are smaller than symbols.}
\end{figure}

\begin{figure}
\centering
\includegraphics[width=0.65\columnwidth]{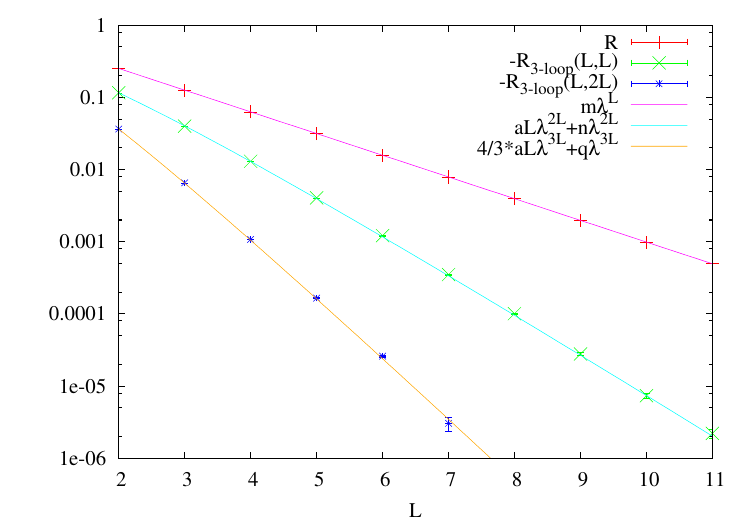}
\caption{\label{Fig:corr}
Response as a function of the distance $L$ on the Bethe lattice, and amputated response on the loop of Fig.~\ref{Fig:loop} with $L_1=L_2=L$. The leading behaviour on a line is $R\propto\lambda^L$, while that on the cubic loop is $R_{3-\text{loop}}\propto L\lambda^{2L}$. Measurements are taken at the critical field $H_c=0.358$ for $z=3$.}
\end{figure}

As shown in Fig.~\ref{Fig:numerical_pL}, both $p_L$ and $c_L$ decay exponentially fast in $L$, as
$$
p_L=a L \lambda^L+b\lambda^L+o(\lambda^L), \quad c_L=c L^2 \lambda^L+dL\lambda^L+e\lambda^L+o(\lambda^L)\;,
$$
where $\lambda$ is the largest eigenvalue of the linear operator associated to the linearization of the BP equations (\ref{eq:BP}) around the fixed point
\begin{equation}
\label{eq:g}
    \lambda\, g(u) = \E_J \int \prod_{i=1}^{z-2} P_B(u_i) du_i\,g(u')\,du'\,\mathbb{I}\left[|H+\sum_i u_i+u'|<|J|\right] \delta\left(u - \sign(J) |H+\sum_i u_i+u'|\right)
\end{equation}
with $\mathbb{I}[.]$ being the indicator function.
At the critical point, $H=H_c$, $\lambda(H_c)=1/(z-1)$ holds.
One can calculate $\lambda$ numerically in the Bethe solution as the growing factor associated
with the evolution of a perturbation.

Once we have a population of triplets at each length $L$, it is quite simple to compute the response function making use of Eq.~(\ref{eq:r-Jh}).

Notice that only events with a non-zero effective coupling contribute to the response function: this is the reason why amplifying the population of cavity messages with $J_L\neq 0$ is mandatory to have a precise measurement of correlations in the $T=0$ limit. In the same way, amplifying the population of cavity messages with $J_L= 0$ and correlated fields will improve the signal for the response function at one loop (see Sec.~\ref{Sec:1loop}).

In Fig.~\ref{Fig:corr}, we show the response function $R^{\text{\tiny{BL}}}(L)$ at distance $L$ averaged over the population of the triplets generated as explained above, in a BL with fixed connectivity $z=3$, at zero temperature and critical field $H_c$: it decays as $R^{\text{\tiny{BL}}}(L) \propto \lambda^L$, with $\lambda=\frac{1}{z-1}$, as already found analytically in the $z\to\infty$ limit \cite{angelini2018one}.

\section{Semi-analytical computation of the distribution of effective triplets on a line}\label{Sec:Ansatz}

In this section, we reproduce the numerical results of the previous section introducing an Ansatz $P_L(u_0,u_L,J)$ for the leading behavior at large $L$ of the joint distribution of the effective coupling and fields between two spins at distance $L$ in a BL at $T=0$.
The Ansatz has the same form of one for the RFIM \cite{angelini2020loop}, except for the fact that the effective coupling $J$ can take both positive and negative values.
The reason why it cannot be different is that in the paramagnetic phase on a tree the RFIM solution at $T=0$ can be mapped into the SG solution in a field through a simple transformation. The Ansatz, at leading order $L\lambda^L$, is the following:
\begin{equation}
\begin{aligned}
P_L(u_0,u_L,J)=& \delta(J)\bigg[ P_B(u_0)P_B(u_L)- b\,L \lambda^L g(u_0)g(u_L)+ \\
& - c_1 L \lambda^L g'(u_0)g'(u_L) - c_2 L \lambda^L g''(u_0)g''(u_L)\bigg]+\\
&+ a L^2 \lambda^L \rho\, e^{-\rho |J| L}g(u_0)g(u_L) 
\end{aligned}
\label{PL}
\end{equation}
Imposing the normalization of the Ansatz, $\int du_0\,du_L\,dJ\,P_L(u_0,u_L,J)\equiv 1$, we obtain the condition $b=2a$. The normalized Ansatz thus becomes:

\begin{equation}
\begin{aligned}
P_L(u_0,u_L,J)=& \delta(J)\bigg[ P_B(u_0)P_B(u_L)- 2a\,L \lambda^L g(u_0)g(u_L)+ \\
& - c_1 L \lambda^L g'(u_0)g'(u_L) - c_2 L \lambda^L g''(u_0)g''(u_L)\bigg]+\\
&+ a L^2 \lambda^L \rho\, e^{-\rho |J| L}g(u_0)g(u_L) 
\end{aligned}
\label{PL_norm}
\end{equation}

At this point, we impose the self-consistency of the part with $J\neq 0$: 
joining two chains of length $L_1$ and $L_2$, the $J\neq 0$ part has to keep the same form with the only substitution of $L$ with $L_1+L_2$.
This is true if the condition
\begin{equation}
a = \frac{\rho}{4 \widehat{P}(0)}
\end{equation} 
holds, where 
\begin{equation}
\widehat P(h) = \int g(u)\,du\,g(v)\,dv\, \prod_{i=1}^{z-2} P_B(u_i)\,du_i\, \delta\left(h-(H+u+v+\sum_i u_i)\right)\;.
\label{eq:Phat}
\end{equation}
We then impose the self-consistency of the whole Ansatz: joining
 on a central spin $\tau$ two chains of length $L_1$ and $L_2$, for which the distribution of fields and coupling is given in eq. (\ref{PL_norm}), following the rules in eq. (\ref{eq:rules}) for the computation of the resulting new triplet of effective fields and coupling once one sums over $\tau$, one should obtain a distribution that has the same form as the one in eq. (\ref{PL_norm})
with $L=L_1+L_2$. This is true only if $c_1=c_2=0$, neglecting contribution  $O(\lambda^L)$, that are already ignored in eq. (\ref{PL_norm}). The final Ansatz thus takes the following form
\begin{equation}
P_L(u_0,u_L,J)= \delta(J)\bigg[ P_B(u_0)P_B(u_L)- 2a\,L \lambda^L g(u_0)g(u_L) \bigg]+a L^2 \lambda^L \rho\, e^{-\rho |J| L}g(u_0)g(u_L)\;.
\label{PLfinal_norm}
\end{equation}
We stress that all the quantities entering this Ansatz can be computed analytically \cite{parisi2014diluted}: $P_B(u)$ from Eq.~(\ref{eq:P}), $g(u)$ from Eq.~(\ref{eq:g}) and $\rho$ from the decay of the mean coupling on branches with $J\neq 0$ or from the properties of the linear operator discussed in detail in the SI of Ref.~\cite{angelini2020loop}.

As for the RFIM, the Ansatz is encoding the fact that at $T=0$ the quantity $J_L$ is either exactly 0 or of order $1/L$ with a probability of order $L\lambda^L$. 
This continuous distribution plus a peak at $J=0$ for the renormalized coupling was already displayed in the finite-dimensional lattice analyzed with the MK RG near the $T=0$ critical point in Ref.~\cite{angelini2017real}.

\section{Computation of the one loop contribution at finite $z$} \label{Sec:1loop}

\begin{figure}
\centering
\includegraphics[width=0.5\columnwidth]{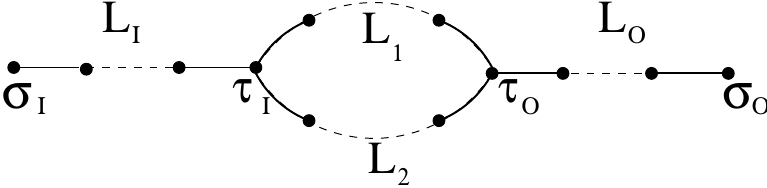}
\caption{\label{Fig:loop}
Topological loop responsible for the leading correction to the critical value of the response function}
\end{figure}

Now we aim at extending the above Ansatz to describe the joint probability distribution of the cavity fields relevant in the computation of the correlation in a loop of the type shown in Fig.~\ref{Fig:loop}.
Some considerations can help us. First of all, the path between the two spins should be connected (i.e.\ made on non-zero effective couplings) if we want a non-zero response function, 
thus on the external legs, the only possible term that we can put is of the form $\lambda^{L_I+L_O}$. 
For this reason, we will just compute what in jargon is called the \textit{amputated} correlation function on the internal branches of the loop.
In the internal loop we have to compute the convolution of two triplets $(J_1,u_i^1,u_o^1)$ and $(J_2,u_i^2,u_o^2)$ of effective fields and coupling at the end of chains of lengths $L_1$ and $L_2$
\begin{multline}
    P_{L_1,L_2}(u_i,u_o,J) =\int du_i^1\, du_i^2\, du_o^1\, du_o^2\, dJ_1\, dJ_2\,  P(u_i^1,u_o^1,J_1)\, P(u_i^2,u_o^2,J_2)\\
    \delta(u_i-u_i^1-u_i^2)\, \delta(u_o-u_o^1-u_o^2)\, \delta(J-J_1-J_2)\;.
    \label{eq:Ploop}
\end{multline}
We have already said that the loop contribution to the observable is the \textit{connected} one, given by the value of the observable computed on the loop minus the observable 
computed on the two paths $L_I+L_1+L_O$ and $L_I+L_2+L_O$ considered as independent. 
We can easily obtain this loop correction disregarding the asymptotic term $P_B(u_I)P_B(u_O)$ in the Ansatz for the two internal branches of length $L_1$ and $L_2$.

There are two relevant contributions to the connected loop: the first one, which we call $P^A$, is the one obtained by a loop in which both $J_1\neq 0$ and $J_2\neq 0$.
The second one, called $P^B$, have $J_1\neq 0$ and $J_2= 0$ or $J_1= 0$ and $J_2\neq 0$. The third contribution with $J_1=J_2=0$ reduces to a disconnected loop, giving no contribution to the connected correlation.
Performing the integral in eq. (\ref{eq:Ploop}), we obtain for the two interesting contributions:
\begin{equation}
\begin{aligned}
P^A_{L_1,L_2}(u_i,u_o,J)&=2\lambda^{L_1+L_2}\widehat{P}(u_i)\widehat{P}(u_o)a^2 \rho L_1^2L_2^2\frac{e^{-|J|L_2\rho}L_1-e^{-|J|L_1\rho}L_2}{L_1^2-L_2^2}\\
P^B_{L_1,L_2}(u_i,u_o,J)&=-2\lambda^{L_1+L_2}\widehat{P}(u_i)\widehat{P}(u_o)a^2 \rho \left(L_1e^{-|J|L_1\rho}+L_2e^{-|J|L_2\rho}\right)
\end{aligned}
\end{equation}
with $\widehat{P}(h)$ defined in eq. (\ref{eq:Phat}).
At this point we average the response over the loop probability distribution obtaining:
\begin{equation}
\begin{aligned}
R^A & \equiv \prob\left[|J|>|u_o|;(u_i,u_o,J)\sim P^A_{L_1,L_2}\right] = 2\,a\,\lambda^{L_1+L_2} \frac{L_1^3-L_2^3}{L_1^2-L_2^2}\\
R^B & \equiv \prob\left[|J|>|u_o|;(u_i,u_o,J)\sim P^B_{L_1,L_2}\right] = -2\,a\,\lambda^{L_1+L_2} \left(L_1+L_2\right)
\end{aligned}
\end{equation}

For the RFIM, $R^A=-R^B$ and the leading contribution to the response coming from the loop is null, while for the SG it is
\begin{equation}
\begin{aligned}
R_{3-\text{loop}}^\text{\tiny BL}(\vec{L})&\equiv\lambda^{L_I+L_O} (R^A+R^B)=-2\,a\,\lambda^{\Sigma(\vec{L})}\frac{L_1L_2}{L_1+L_2}\\
\label{eq:Rloop}
\end{aligned}
\end{equation}
with $\Sigma(\vec{L})=L_I+L_1+L_2+L_O$.
The two branches act as resistors in parallel.
We can check this result numerically by computing the amputated one-loop response using random triplets for the two branches obtained from the enriched populations discussed above.
Putting $L_1=L_2=L$ in Eq.~(\ref{eq:Rloop}), we have a leading behaviour of the type $R_{3-\text{loop}}(L,L)=aL\lambda^{2L}$, that perfectly describes the data in Fig.~\ref{Fig:corr}.
To check the dependence on $L_1$ and $L_2$, we 
numerically verified that putting $L_1=2L_2=2L$ one obtains
$R_{3-\text{loop}}(L,2L)=-\frac{4}{3} a L \lambda^{3L}$, as shown in Fig.~\ref{Fig:corr}.

\section{The $T=0$ limit of the solution obtained with $T > 0$ and $z\to\infty$}

In Ref.~\cite{angelini2018one} we computed at positive temperatures ($T>0$) the analytical expression for the connected correlation function on a BL with a manually inserted cubic loop in the $z\to\infty$ limit (to recover previous results obtained in the fully-connected model). The result can be summarized as follows:
\begin{align}
\nonumber
R_{3-\text{loop}}^{z=\infty}=&\lambda_R^{L_I+L_O}\left[\left(L_2 \lambda_{L/A}^{L_2-1} \lambda_R^{L_1}+L_1 \lambda_{L/A}^{L_1-1} \lambda_R^{L_2}\right)b_1 \beta^2
+\lambda_{L/A}^{L_1+L_2}b_2 \right.\\
&\left.+(\lambda_{L/A}^{L_2} \lambda_R^{L_1}+\lambda_{L/A}^{L_1} \lambda_R^{L_2})b_3+\lambda_R^{L_1+L_2} b_4\right]
\label{Eq:CCloop}
\end{align}
with $\lambda_R$ the eigenvalue associated to the replicon sector, $\lambda_{L/A}$ the degenerate longitudinal-anomalous eigenvalues and $\beta=1/T$ the inverse temperature.
The coefficients read
\begin{align}
\nonumber b_1=&-32 (2 m_2-3 m_4) (1-7 m_2+11 m_4-5 m_6)^2 \\
\nonumber b_2=&64 (1-7 m_2+11 m_4-m_6)^2 \\
\nonumber b_3=&-80 (1+35 m_2^2+77 m_4^2+m_4 (18-68 m_6)-2 m_2 (6+52 m_4-23m_6)-8 m_6+15 m_6^2)\\
\label{eq:coeff_b} b_4=&32 \left(1+44 m_2^2+101 m_4^2+m_4 (22-90 m_6)-2 m_2 (7+67 m_4-30 m_6)-10 m_6+20 m_6^2\right) 
\end{align}
with $m_2$, $m_4$ and $m_6$ the moments of order 2, 4, 6 of the magnetization. Their definition is the following
\begin{equation}
m_a=\frac{1}{\sqrt{2 \pi}}\int^{\infty}_{-\infty} e^{-z^2/2} \tanh^a\left(\beta(\sqrt{Q} z+H)\right)dz\;,
\label{eq:m_gaus}
\end{equation}
with $Q$ solution of the self-consistency equation $Q=m_2$.
One thus could look at the $T=0$ limit of these expressions and compare them to what we have presented above.

Let us emphasize that for any positive temperature on the dAT line $\lambda_R$ is the only critical eigenvalue and thus the leading term in the response function is of the type $\lambda_R^{\Sigma(\vec{L})}$. 
At $T=0$, however, $\lambda_{L/A}\to\lambda_R$ and the leading terms in the one-loop response could change.

To compute the $T=0$ limit, we should substitute in the equations the values for the moments of the magnetization in the zero temperature limit.
Expanding Eq.~(\ref{eq:m_gaus}) around $T=0$ we find
\begin{align}
\nonumber
m_a=& 1 - C_1(a) \left[\frac{T e^{-H^2/2}}{\sqrt{2\pi}} - \left(\frac{T e^{-H^2/2}}{\sqrt{2\pi}}\right)^2 (H^2-1) + \left(\frac{T e^{-H^2/2}}{\sqrt{2\pi}}\right)^3 \frac{3 H^4 - 10 H^2 + 5}{2} \right] +\\
& - C_2(a) \frac{T^3 e^{-H^2/2}}{2\sqrt{2 \pi}} (H^2 - 1)\;,
\label{eq:maExp}
\end{align}
with $C_1(a)\equiv\int^{\infty}_{-\infty} (1-\tanh^a(x))dx$ and $C_2(a)\equiv\int^{\infty}_{-\infty} x^2(1-\tanh^a(x))dx$, whose numerical values are $C_1(2)=2$, $C_1(4)=8/3$, $C_1(6)=46/15$, $C_2(2)=\pi^2/6$, $C_2(4)=2(3+\pi^2)/9$, $C_2(6)=4/3+23\,\pi^2/90$.

Substituting expression (\ref{eq:maExp}) in the coefficients defined in Eq.~(\ref{eq:coeff_b}) we find $b_1=O(T^6)$, $b_2=O(T^6)$, $b_3=O(T^4)$ and $b_4=O(T^2)$, implying that in the $T=0$ limit the dominating term for the response function is the one coming from the replicon, as for $T\neq0$, that has a behaviour $\lambda_R^{\Sigma(\vec{L})}$. 
This result is different from what we computed directly at $T=0$ and finite $z$.

One may attempt to improve this result, obtained in the $z\to\infty$ limit, by using the actual values of the magnetization moments computed at finite $z$ in the Bethe solution. This would represent the simplest idea to go beyond the $z=\infty$ result.
However, we find that the coefficients entering the expansions up to the second order in $T$ of the magnetization moments,
\begin{eqnarray}
m_2&=&1+m_2^{(1)}T+m_2^{(2)}T^2+O(T^3)\;,\\
m_4&=&1+m_4^{(1)}T+m_4^{(2)}T^2+O(T^3)\;,\\
m_6&=&1+m_6^{(1)}T+m_6^{(2)}T^2+O(T^3)\;,
\end{eqnarray}
do satisfy the following relations
\begin{equation}
\frac{m_4^{(1)}}{m_2^{(1)}}=\frac{m_4^{(2)}}{m_2^{(2)}}=\frac{4}{3}\qquad
\frac{m_6^{(1)}}{m_2^{(1)}}=\frac{m_6^{(2)}}{m_2^{(2)}}=\frac{23}{15}\;,
\end{equation}
independently on the distribution of the cavity fields. This implies that using the non-Gaussian cavity fields obtained at finite $z$ would not change the result obtained in the $z\to\infty$ limit, that is using Gaussian fields. This is true up to second order in $T$, and this is enough as the leading term $b_4$ is of that order of magnitude. We checked this result numerically for a BL with a finite and small value for $z$.

The results of this Section prove that computing the leading term of the correlation at $T>0$, where the only critical eigenvalue is the replicon, inevitably produces a wrong result in the $T\to 0$ limit, where the degeneracy among different eigenvalue plays an important role. This is true both in the fully connected model ($z\to\infty$) and also for $z$ finite within the Bethe solution.
The only way to perform the right computation is by working directly at $T=0$ as we have done in the present work.

\end{document}